\documentclass[pra,twocolumn,superscriptaddress,10pt,showpacs,floatfix]{revtex4-1}
\usepackage[pdftex,plainpages=false,colorlinks=true,citecolor=blue,linkcolor=blue,urlcolor=blue,filecolor=green,bookmarksopen=true]{hyperref}
\usepackage[utf8]{inputenc}
\usepackage{amsmath}
\usepackage[caption = false]{subfig}
\usepackage{graphicx,epstopdf}
\usepackage{blindtext}
\usepackage{lipsum}
\usepackage{amsfonts}
\usepackage{bbm}
\usepackage{amssymb}
\usepackage{enumerate}
\usepackage{color}
\usepackage{latexsym}
\usepackage{times,txfonts}

\newcommand{\Fcal}{\mathcal{F}}

\newcommand{\ket}[1]{| #1 \rangle}

\newcommand{\interpro}[2]{\langle #1 | #2 \rangle}
\newcommand{\bra}[1]{\langle #1 |}

\begin{document}

\title{Minimal resource to design spin-based quantum transistors}

\author{Alan C. Santos}
\email{ac\_santos@id.uff.br}
\affiliation{Instituto de F\'{i}sica, Universidade Federal Fluminense, Av. Gal. Milton Tavares de Souza s/n, Gragoat\'{a}, 24210-346 Niter\'{o}i, Rio de Janeiro, Brazil}

\begin{abstract}
Designing quantum analogous of classical computers components is the heart of quantum information processors. In this sense, for quantum devices, quantum transistors are believed to be as necessary as the classical ones for classical devices. In this paper we design the smallest spin-based quantum transistor. In fact, while previous schemes explore entangled quantum state for simulating the performance of quantum transistors gate (open and close it), in this paper we show that such task can be achieved by a controllable external magnetic field in a three-spin quantum system. Thus, we could reduce the number of physical spins required to design the quantum transistor, since the gate in our transistor is composed by a single-spin, instead two-spin systems. To analyze the performance of our quantum transistor, we consider its robustness against two decohering environments.
\end{abstract}

\maketitle

\section{Introduction}

The miniaturization of classical computers components to quantum regime is the challenger for building quantum technologies, like quantum processors that have been built using superconducting qubits \cite{Johnson:11,Barends:16,Orlando:99,Harris:10} or trapped ions \cite{Monroe:13,Kielpinski:02,Gulde:03,Brown:16}, for example. Such miniaturization of classical components has an strong relationship with the well-known Moore's law, that ``establishes" (predicts) that number of transistor doubles approximately every two years \cite{Moore:98}. In this sense, inspired by important role of classical transistor, building transistors that works in quantum scale has awake the attention in last decade \cite{Devoret:98,Igor:04,Marchukov:16}. Quantum transistors have been studied in scenario of adiabatic quantum computation \cite{Williamson:15,Bacon:17}, ultra-cold atoms \cite{Micheli:04,Vaishnav:08,Fuechsle:12}, spin chain \cite{Marchukov:16} and among others physical systems \cite{Gajdacz:14,Chang:07,Gardelis:99,Chen:13,Hwang:09,Bose:12}.

As it was discussed in Refs. \cite{Marchukov:16,Loft:18}, we can efficiently use an spin chain to simulate a transistor in quantum regime. The smallest transistor, as proposed in Refs. \cite{Marchukov:16,Loft:18}, is composed by four qubits separated into three components: \textit{source} (one qubit), \textit{drain} (one qubit) and \textit{gate} (two qubits). The transistor gate in such transistor is simulated by creating an entangled quantum state between two spins (closed gate), while we can open the gate if the system is a non-entangled state. In this sense, the gate control requires our ability for manipulating internal degree of freedom of the system. In this paper we provide an alternative spin based quantum transistor, where no manipulation of internal degree of freedom is required, consequently an entangled quantum state is not necessary for simulating the transistor gate.

Based on Moore's predictions \cite{Moore:98}, we will show how we can save one qubit by designing a three-qubit-spin based quantum transistor. While the gate control in four-spin-based quantum transistor is achieved by controlling an two-qubit entangled state, our protocols allow us to save one qubit and no entangled quantum control is required. To this end, the gate control is performed by adjusting the magnetic field that acts on the gate qubit. To end, we study the robustness of our system against decohering effects.

\section{A quantum transistor with three spins}

\begin{figure}[t!]
	\centering
	\includegraphics[scale=0.2]{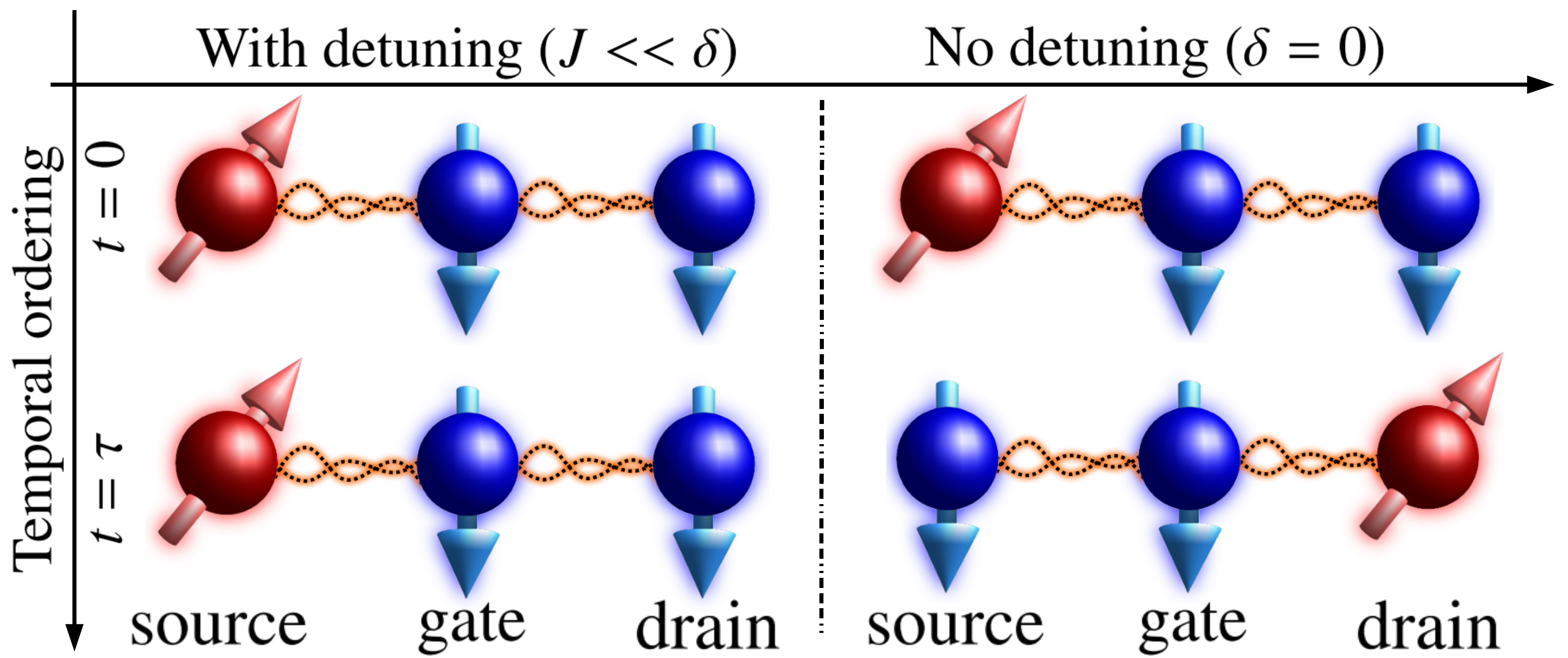}
	\caption{Pictorial representation of a three-spin-based quantum transistor. By applying a detuning $\delta \gg J$ on the gate (meddle qubit), where $J$ is the strength of the spin-spin interaction, the quantum information is kept in source qubit (left qubit). When we turn off the detuning ($\delta = 0$) the quantum information will flow from source to drain (right qubit)} \label{Scheme}
\end{figure}

The spin-based quantum transistor presented in this letter is composed as shown in Fig. \ref{Scheme}. The quantum information is prepared (encoded) on the source qubit and it can flow from source to drain if the gate is open, otherwise the information is kept in source qubit. 

In order to show how our transistor works, let us consider the XY spin model with three qubits in presence of a $Z$-directional static magnetic fields. The Hamiltonian reads as
\begin{eqnarray}
H = \hbar \sum_{i=1}^{3} \omega_{i} \sigma^{(i)}_{z} + \frac{\hbar J}{2}\sum_{i=1}^{2}\left[\sigma^{(i)}_{x}\sigma^{(i+1)}_{x} + \sigma^{(i)}_{y}\sigma^{(i+1)}_{y}\right] \mathrm{ , } \label{H}
\end{eqnarray}
where we will consider $\omega_{1} = \omega_{3} = \omega_{0}$ and $\omega_{2} = \omega_{0} + \delta$, where the detuning $\delta$ concerning $\omega_{1}$ and $\omega_{3}$ will plays an important role in our study. The input state of the system is considered as $\ket{\Psi(0)} = \ket{\psi}_{\mathrm{s}}\ket{\downarrow}_{\mathrm{g}}\ket{\downarrow}_{\mathrm{d}}$ (as we show in the Fig. \ref{Scheme}), where $\ket{\psi}_{\mathrm{s}} = \alpha \ket{\uparrow}_{\mathrm{s}}+\beta\ket{\downarrow}_{\mathrm{s}}$. 

Thus, by letting the system evolves by Hamiltonian $H$, we get the evolved state $\ket{\Psi(t)}$ given by
\begin{eqnarray}
\ket{\Psi(t)} = U(t)\ket{\Psi(0)} = \exp \left( -\frac{i}{\hbar} Ht \right) \ket{\Psi(0)} \mathrm{ , }
\end{eqnarray}
so that we can compute the time-dependent probability $p(t)$ of finding the system in state $\ket{\Psi(0)}$. To design a quantum transistor we need to show how our system can block the transfer quantum information from source to drain. To compute the blockade probability $p(t) = |\interpro{\Psi(0)}{\Psi(t)}|^2$, let us use the Hamiltonian's symmetry. Once the Hamiltonian in Eq. \eqref{H} preserves the magnetization, it is possible to show that 
\begin{eqnarray}
U(t)\ket{\downarrow}_{\mathrm{s}}\ket{\downarrow}_{\mathrm{g}}\ket{\downarrow}_{\mathrm{d}} = e^{i\theta(t)}\ket{\downarrow}_{\mathrm{s}}\ket{\downarrow}_{\mathrm{g}}\ket{\downarrow}_{\mathrm{d}} \mathrm{ , }
\end{eqnarray}
for some real parameter $\theta(t)$. Therefore, it is possible to write $p(t) = |\alpha|^2 + p_{\uparrow \downarrow \downarrow}(t)|\beta|^2$, where
\begin{eqnarray}
p_{\uparrow \downarrow \downarrow}(t) = 
|\bra{\uparrow}_{\mathrm{s}}\bra{\downarrow}_{\mathrm{g}}\bra{\downarrow}_{\mathrm{d}}U(t)\ket{\uparrow}_{\mathrm{s}}\ket{\downarrow}_{\mathrm{g}}\ket{\downarrow}_{\mathrm{d}}|^2 \text{ . }
\end{eqnarray}

We can see that for obtaining $p(t) = 1$, we need to get $p_{\uparrow \downarrow \downarrow}(t)$, because $|\alpha|^2 + |\beta|^2 = 1$. In this case our study is reduced to study $p_{\uparrow \downarrow \downarrow}(t)$, where we can analytically compute $p_{\uparrow \downarrow \downarrow}(t)$ as
\begin{eqnarray}
p_{\uparrow \downarrow \downarrow}(t) &=& \frac{1}{8} \left(3+ \frac{\delta^2}{\Delta^2}\right) + \frac{1}{2} \cos \delta t \cos \Delta t + \frac{J^2}{4\Delta^2} \cos 2 \Delta t \nonumber \\
&+& \frac{\delta}{2\Delta} \sin \delta t \sin \Delta t \label{p}
\end{eqnarray}
where we defined $\Delta^2 = \delta^2+2J^2$. As expected, the probability $p_{\uparrow \downarrow \downarrow}(t)$ is dependent on the interaction parameter $J$ and detuning $\delta$, so that we can adequately adjust them in order to control $p_{\uparrow \downarrow \downarrow}(t)$. If we are interested to provide an optimally controllable quantum transistor, it is reasonable to keep $J$ fixed, while we can manipulate $\delta$. In fact, in physical real experimental scenario $J$ is related with internal degree of freedom of the system \cite{Jones:01,Linden:98,Bernardes:16,Silva:16}, therefore it is simpler to control $\delta$ than $J$.

\emph{Closing the gate -- } As discussed previously, the quantum transistor proposed here is based on our ability to control the detuning parameter $\delta$, associated with the magnetic field along $Z$-axes. Indeed, as we can see from Eq. \eqref{p}, the detuning $\delta$ help us to control the information $\ket{\psi}$ in our system. For example, to block the information let us consider an situation where we have $J<<\delta$. In general, the magnetic fields in $Z$-direction is stronger than coupling between the spins \cite{Sarthour:Book}, therefore it is reasonable to think about this consideration. Thus, in this regime $J<<\delta$ we can expand the $p_{\uparrow \downarrow \downarrow}(t)$ as
\begin{eqnarray}
p^{J<<\delta}_{\uparrow \downarrow \downarrow}(t) &\approx& - \frac{J^4}{8\delta^4} 
[-7+2 \delta^2 t^2 +7 \cos 2\delta t + 6 \delta t \sin 2\delta t] \nonumber \\ &-& \frac{J^2}{\delta^2} \sin^2 \delta t + 1 \mathrm{ . }
\end{eqnarray}

\begin{figure}[t!]
	\centering
	\includegraphics[scale=0.27]{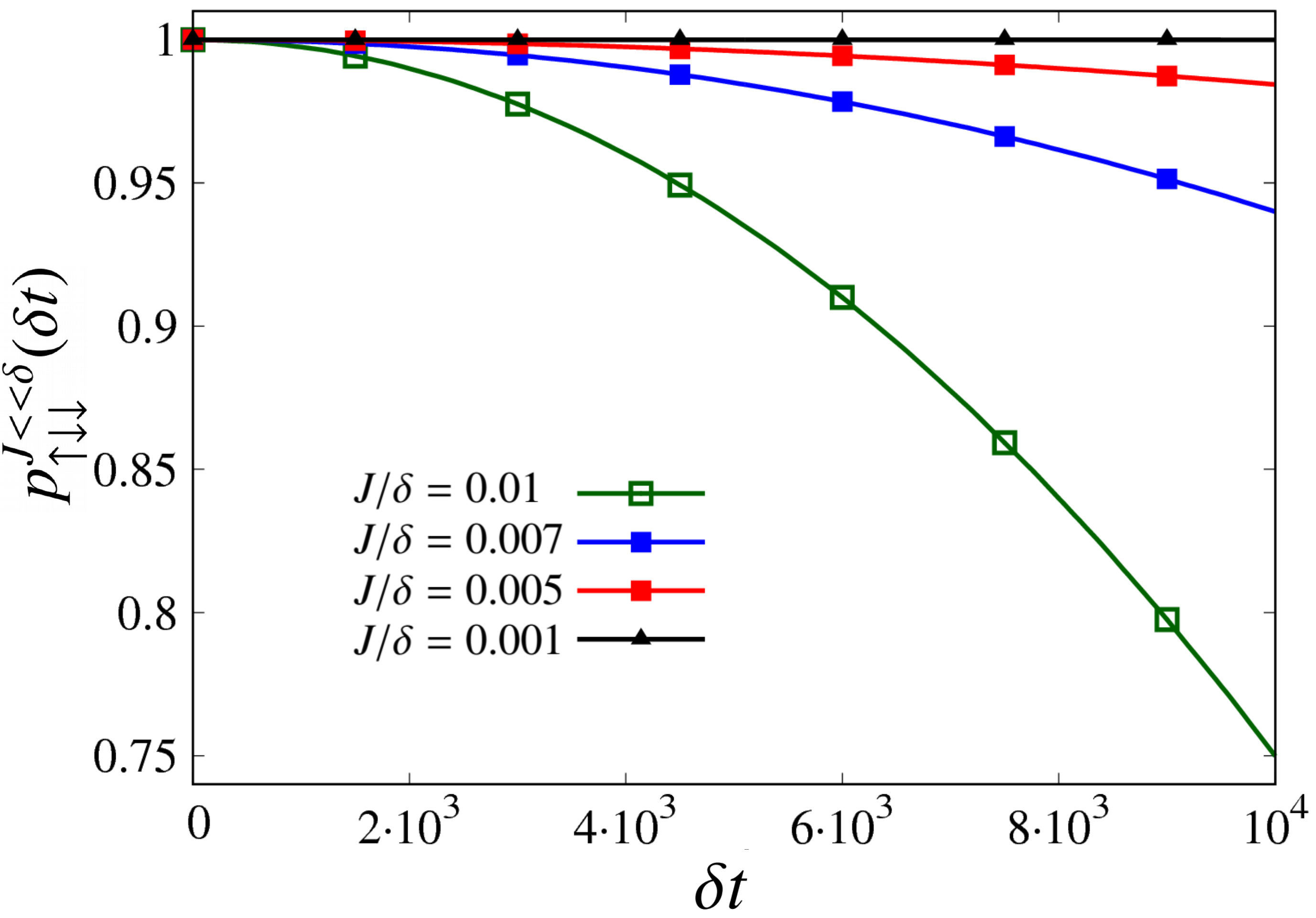}
	\caption{Probability $p^{J<<\delta}_{\uparrow \downarrow \downarrow}$ as function of the dimensionless quantity $\delta t$ for some choices of the quantity $J/\delta$.} \label{Fig1}
\end{figure}

In Fig. \ref{Fig1} we show the behavior of $p^{J<<\delta}_{\uparrow \downarrow \downarrow}(t)$ for some choices of the relation as $J/\delta$. The high fidelity of $p^{J<<\delta}_{\uparrow \downarrow \downarrow}(t)$ allow us to guarantee that the quantum information remains in the source spin for sufficiently long times ($\delta t >> 1$) dependent on the relation $J/\delta$. To provide an experimental concrete analysis, let us consider that $J \sim 1$KHz, when $\delta \sim 1$MHz we have a high fidelity gate closed for values of $t$ as $t \sim 10$ms. As we shall show coming soon, it is a reasonable blockade time.

\emph{Opening the gate -- } Once we can block the quantum information flowing from source to drain adjusting the detuning, it would be important to show how we can use the same parameter to open the gate. Again, we start from Eq. \eqref{p}, but considering a regime where we have no detuning, i.e., putting $\delta=0$ in Eq. \eqref{p}. Under this consideration the Eq. \eqref{p} provide us
\begin{eqnarray}
p^{\delta=0}_{\uparrow \downarrow \downarrow}(t) = \cos^4 \left( \frac{Jt}{\sqrt{2}} \right) \mathrm{ . }
\end{eqnarray}

Thus, we can see that when $Jt = \pi/\sqrt{2}$ we have $p^{\delta=0}_{\uparrow \downarrow \downarrow}(t)=0$, therefore the information is not in source. But, where the information is? To answer this question we need to compute $p^{\delta=0}_{\downarrow \downarrow \uparrow}(t)$, that reads as
\begin{eqnarray}
p^{\delta=0}_{\downarrow \downarrow \uparrow}(t) = \sin^4 \left( \frac{Jt}{\sqrt{2}} \right) \mathrm{ . }
\end{eqnarray}

Let us take account the information that the transfer probability only depend on the coupling strength in this case (as expected, once the parameters $\omega_{i}$ are identical). So, by putting $Jt = \pi/\sqrt{2}$ we find $p^{\delta=0}_{\downarrow \downarrow \uparrow}(t) = 1$. This result shows that the quantum information is transfered to drain at transference time $\tau_{\mathrm{T}} = \pi/J\sqrt{2}$. It is important to note that from $\tau_{\mathrm{T}}$ we can estimate the characteristic blokate time $\tau_{\mathrm{B}}$ of our transistor. In fact, if we consider $J \sim 1$KHz we have $\tau_{\mathrm{T}} \sim 1$ms, while the blocking time for a detuning $\delta \sim 1$MHz is $\tau_{\mathrm{B}} \sim 10\tau_{\mathrm{T}}$.

\section{Robustness against decoherence analysis}

In spin quantum systems we have inevitable decoherence effects due inhomogeneous magnetic fields and many others factors (systematic errors). Therefore, in this section we study the performance of a three-spin-based quantum transistor against such decohering effects. In particular we will analyze the robustness of the transistor against two kind of decoherence effects.

\begin{figure*}[t!]
	\centering
	\subfloat[Transfer fidelity from Eq. \eqref{Lind}]{\includegraphics[scale=0.16]{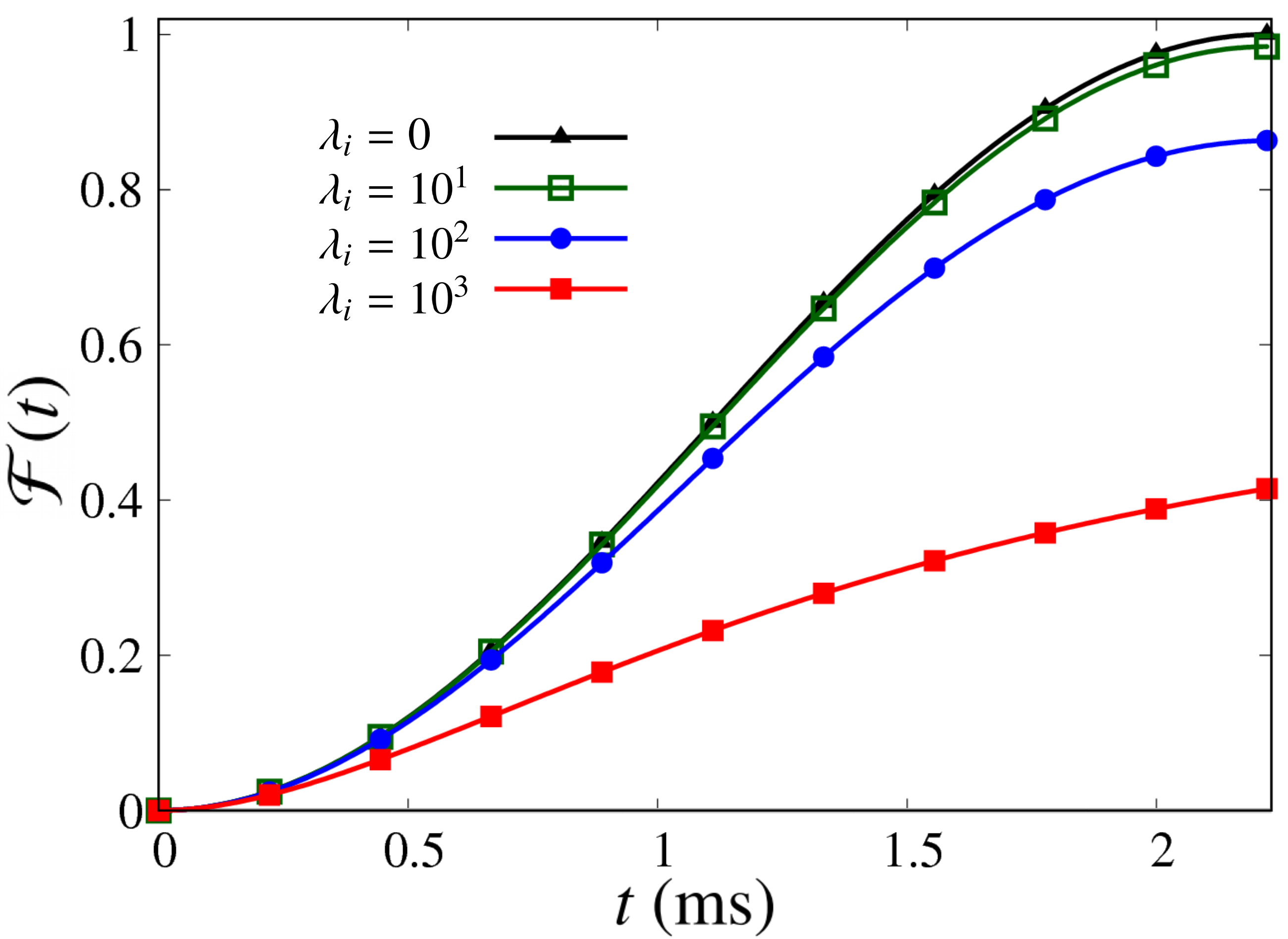}\label{Trans1}} \hspace{1mm}
	\subfloat[Blockade fidelity from Eq. \eqref{Lind}]{\includegraphics[scale=0.165]{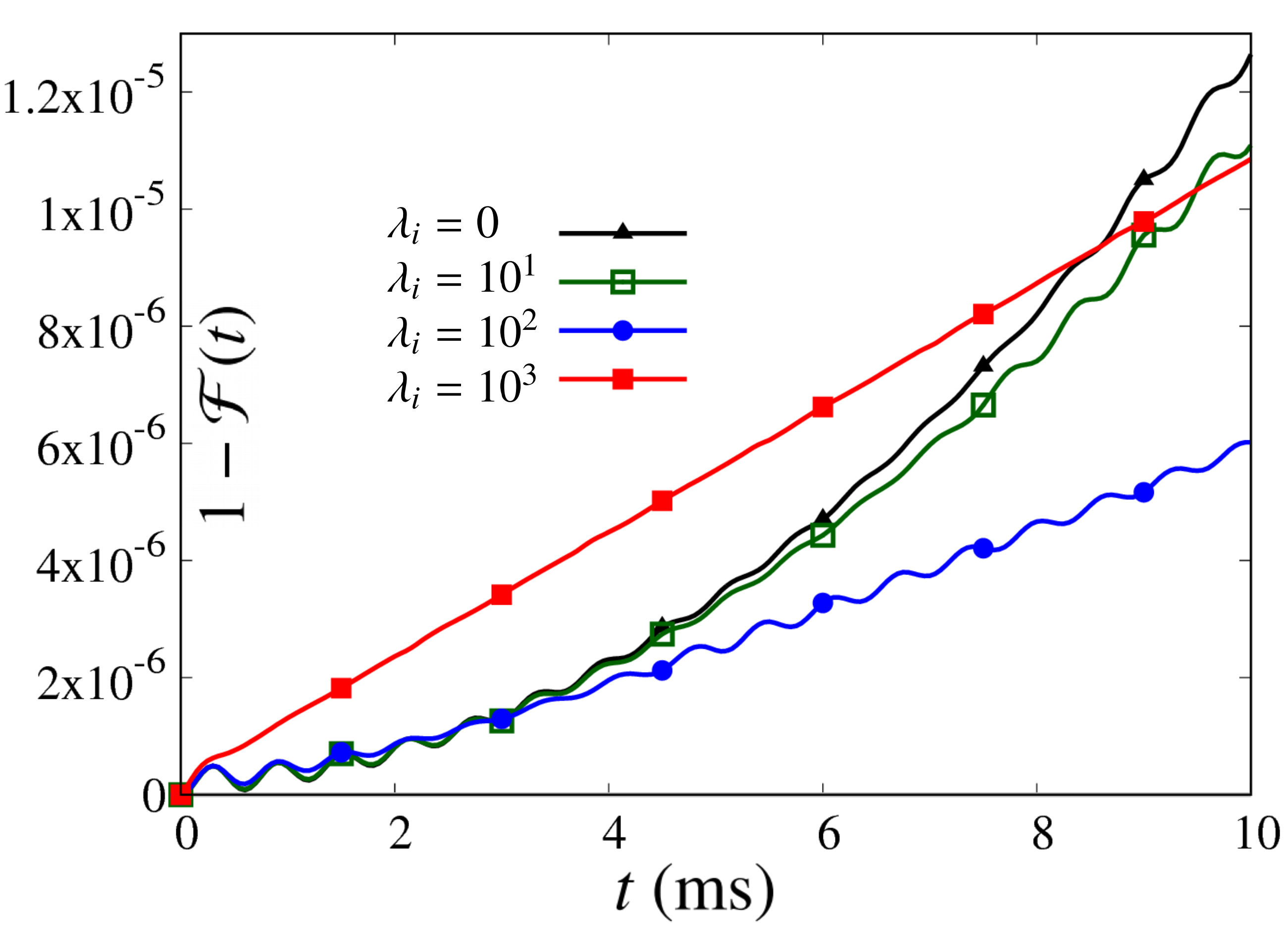}\label{Block1}}\hspace{1mm}
	\subfloat[Transfer fidelity from Eq. \eqref{Milburn}]{\includegraphics[scale=0.16]{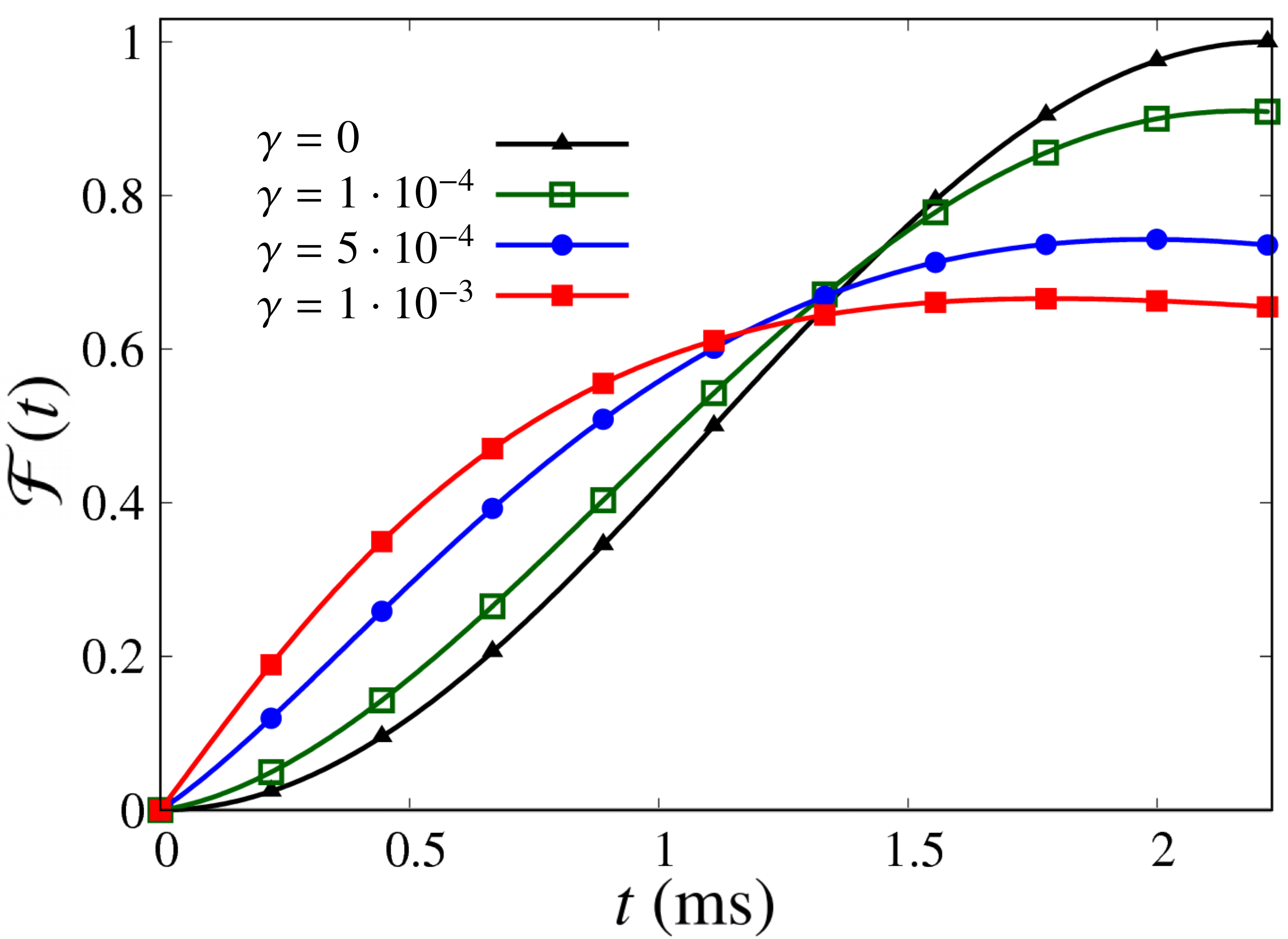}\label{Trans2}}\hspace{1mm}
	\subfloat[Blockade fidelity from Eq. \eqref{Milburn}]{\includegraphics[scale=0.165]{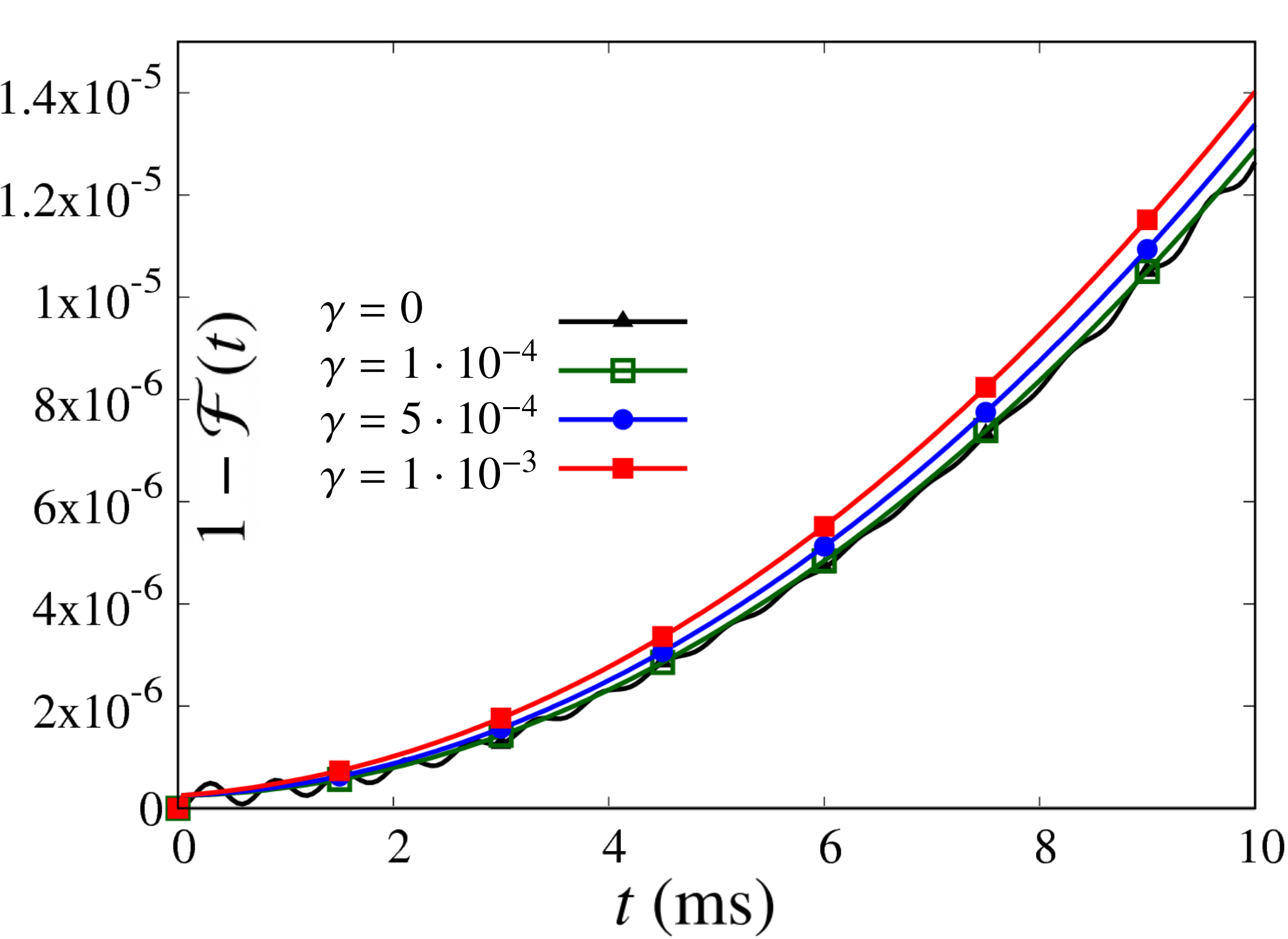}\label{Block2}}
	\caption{Transfer and blockade fidelities obtained from Eqs. \eqref{Lind} and \eqref{Milburn}. For numerical calculations we have considered $\delta = 10^6$Hz (for blockade situations), while $J$ is kept as $J = 10^3$Hz for all of graphs.} \label{Fig3}
\end{figure*}

The first decohering model is based on an weak coupling regime, where the decohering effect acts on a particular spin independently from each others. In this case the dynamics of the system is driven by Lindblad equation \cite{Lindblad:76}
\begin{eqnarray}
\dot{\rho} (t) = \frac{1}{i \hbar} [H,\rho(t)] + \sum_{i=1}^{3} \lambda_{i} \left[\sigma_{z}^{i}\rho(t)\sigma_{z}^{i} - \rho(t)\right] \mathrm{ , } \label{Lind}
\end{eqnarray}
with $\lambda_{i}$ the dephasing rate of the $i$-th spin (here we will consider identical dephasing rate). The second decohering error considered here is associated with the so called intrinsic decohering effects in quantum systems, as proposed by the Milburn’s intrinsic decoherence theory \cite{Milburn:91}. The master equation is given by
\begin{eqnarray}
\dot{\rho} (t) = \frac{1}{i \hbar} [H,\rho(t)] - \frac{\gamma}{2} [H,[H,\rho(t)]]  \mathrm{ , } \label{Milburn}
\end{eqnarray}
where $\gamma$ is the decohering decoherence rate. The Milburn’s model has been considered to study intrinsic decoherence in optical systems \cite{Moya-Cessa:93,Schneider:98,Plenio:98}, spin channels \cite{Hu:09}, among others \cite{Levy:18,Santos:18-a,X-Jing:13,Zheng:17}. 

In order to quantify the robustness of our protocol, we will consider the Bures metric for pure quantum states given by \cite{Bures:69}
\begin{eqnarray}
\Fcal (t) = \sqrt{\bra{\psi_{\mathrm{tar}}}\rho(t)\ket{\psi_{\mathrm{tar}}}} \mathrm{ , }
\end{eqnarray}
where $\ket{\psi_{\mathrm{tar}}}$ is the target state, in our case $\ket{\psi_{\mathrm{tar}}} = \ket{\psi}_{\mathrm{s}}\ket{\downarrow}_{\mathrm{g}}\ket{\downarrow}_{\mathrm{d}}$, and $\rho(t)$ is solution of the non-unitary process. This metric has been considered in several applications in quantum information, like quantum speed limit \cite{DeffnerPRL:13,Deffner:13}, studies on quantum technologies \cite{Paris:09}, quantum phase transitions \cite{Gu:10} and others applications \cite{Santos:18-b,Hill:97,Braunstein:94,Nielsen:Book}. The results for Lindblad equation and Milburn equation are shown in Figs. \ref{Fig3}.

As one can see from Figs. \ref{Fig3}, the performance of a three-spin-based quantum transistor is found to be good against some decohering effects, but naturally there are others non-unitary processes where such performance is not good. Remarkably, we have high fidelity of blockade process for the two decohering effects considered here (see Figs. \ref{Block1} and \ref{Block2}), while the transfer fidelity is not so good as the blockade one for the same decohering regime (see Figs. \ref{Trans1} and \ref{Trans2}). In general, this asymmetric behavior between blockade and transfer fidelities is obtained in others quantum transistors \cite{Marchukov:16,Loft:18,Vaishnav:08}.

In general, the Hamiltonian presented in Eq. \eqref{H} is not a particular Hamiltonian of spin quantum systems. If we can simulate  Heisenberg-XY interactions qubit in Eq. \eqref{H}, we can use this protocol for designing an three-qubit quantum transistor. For instance, this task can be achieved from different experimental scenarios, like nuclear magnetic resonance experiments \cite{Zhang:05}, cold atoms \cite{Murmann:15} and trapped ions \cite{Grass:14,Cui:16}.

\section{Conclusions}

A three-spin-based quantum transistor is proposed in this letter. Once an spin-based quantum transistor is composed by a \textit{source}, \textit{gate} and \textit{drain} qubit, the smallest quantum transistor should be composed by three spins, as proposed in this section. For this reason, we believe that our scheme could be view as alternative to spin-based quantum transistor in situations where very small quantum systems are necessaries. Obviously, optimal control of magnetic fields is a requisite to design three-spin quantum transistor. Thus, our model would be recommended in situations where the magnetic fields control is simpler than entangled state control.

As a principle proof, our three-spin system could be efficiently implemented in Nuclear Magnetic Resonance (NMR). NMR molecules used to implement quantum protocols with three-qubit systems \cite{Jones:01,Linden:98,Bernardes:16,Silva:16} would be a candidate for enconding and engineering our quantum transistor. Actually, an experimental study of robustness against decoherence effects of our model is an open question to be considered in future research.

\begin{acknowledgments}
This work was supported by national agencies CNPq-Brazil and National Institute for Science and Technology of Quantum Information (INCT-IQ). I would like to thank to professor Dr. Marcelo S. Sarandy, from Federal Fluminense University, for his support and for encouraging me to follow this research.
\end{acknowledgments}

\end{document}